\documentclass[twocolumn,aps,prl,superscriptaddress,reprint]{revtex4-2}
\usepackage{textcomp}
\usepackage[utf8]{inputenc}
\usepackage{amsmath}
\usepackage{cancel}
\usepackage{graphicx}
\usepackage{titlesec}
\titlespacing\section{0pt}{7pt plus 3pt minus 2pt}{2pt plus 1pt minus 1pt}

\makeatletter

\usepackage{amsfonts}

\usepackage[colorinlistoftodos,prependcaption,textsize=small]{todonotes}
\usepackage{color}
\definecolor{red}{rgb}{0.75,0,0}
\definecolor{blue}{rgb}{0,0,0.75}
\definecolor{green}{rgb}{0,0.5,0}
\definecolor{orange-red}{RGB}{255,69,0}

\newcommand{\eq}{\begin{equation}}
\newcommand{\eeq}{\end{equation}}
\newcommand{\aeq}{\begin{equation}\begin{aligned}}
\newcommand{\eaeq}{\end{aligned}\end{equation}}

\makeatother

\begin{document}
\title{{Bacterial turbulence} drives interfacial waves and shape dynamics in phase-separated droplets}
\author{Kan Chang}
\thanks{These authors contributed equally}
\affiliation{School of Physics and Astronomy, Institute of Natural Sciences and
MOE-LSC, Shanghai Jiao Tong University, Shanghai 200240, China}
\author{Yulin Li}
\thanks{These authors contributed equally}
\affiliation{Fujian Provincial Key Laboratory for Soft Functional Materials Research,
Research Institute for Biomimetics and Soft Matter, Department of
Physics, Xiamen University, Xiamen, Fujian 361005, China}
\author{Ming Yuan}
\affiliation{School of Physics and Astronomy, Institute of Natural Sciences and
MOE-LSC, Shanghai Jiao Tong University, Shanghai 200240, China}
\author{Masaki Sano}
\affiliation{School of Physics and Astronomy, Institute of Natural Sciences and
MOE-LSC, Shanghai Jiao Tong University, Shanghai 200240, China}
\affiliation{Universal Biology Institute, Graduate School of Science, The University of Tokyo, Tokyo 113-0033, Japan}

\author{Zhihong You}
\thanks{Corresponding author: zhyou@xmu.edu.cn}
\affiliation{Fujian Provincial Key Laboratory for Soft Functional Materials Research,
Research Institute for Biomimetics and Soft Matter, Department of
Physics, Xiamen University, Xiamen, Fujian 361005, China}
\author{H. P. Zhang}
\thanks{Corresponding author: Hepeng\_zhang@sjtu.edu.cn}
\affiliation{School of Physics and Astronomy, Institute of Natural Sciences and
MOE-LSC, Shanghai Jiao Tong University, Shanghai 200240, China}
\date{\today}

\begin{abstract}
Liquid–liquid phase separation is important across biology, physics, and materials science. Although usually studied at equilibrium, active components—such as motor proteins, enzymes, and synthetic microswimmers—are increasingly recognized as key players in reshaping phase separation dynamics.
{Yet how internally generated active stresses are transmitted to capillary interfaces to reshape three-dimensional droplet dynamics remains poorly understood.} Here, we encapsulate dense suspensions of motile bacteria inside phase-separated aqueous droplets, creating a closed droplet whose interface is driven from within by {bacterial turbulence}. By varying bacterial density, we control the active stress at the droplet interface. {At low bacterial density}, we observe scale-dependent interfacial fluctuations that propagate as waves. In this low Reynolds number regime, these waves arise from an effective inertial response, generated when active bacterial stresses balance passive viscous damping of the interface. {At higher bacterial density}, droplets deform strongly—exceeding the Plateau–Rayleigh instability threshold—and even form bacteria-scale filaments—a morphology without a passive counterpart. Enhanced droplet motility and accelerated coarsening accompany these shape changes. Our work shows how active stresses can reshape the morphology and dynamics of multiphase systems, offering new insight into the physics of internally driven phase-separated fluids.
\end{abstract}
\maketitle


\vspace{-0.5em}
\section*{Introduction}

Interfaces play a central role in living matter. From cell membranes to biomolecular condensates formed by liquid–liquid phase separation (LLPS),  these boundaries separate biochemical reactions and regulate material exchange with the environment~\cite{needleman_active_2017,LvLi2025_InterfaceMorphodynamics,TurlierBetz2019_ActiveLivingMembraneFluctuations,HymanWeberJuelicher2014_LLPhaseSeparation,FletcherMullins2010_CellMechanicsCytoskeleton,PollardCooper2009,KerenPincusAllenBarnhartMarriottMogilnerTheriot2008,BrochardLennon1975,GovZilmanSafran2003,FournierLacosteRaphael2004,GovSafran2005,TurlierFedosovAudolyAuthGovSykesJoannyGompperBetz2016,Schwayer2019_ZO1,KimKalappurakkalMayorRosen2019,WiegandHyman2020}. To sustain biological functions, such interfaces are continually driven out of equilibrium by active forces. For example, actin networks power shape changes and cell motility~\cite{FletcherMullins2010_CellMechanicsCytoskeleton,PollardCooper2009,KerenPincusAllenBarnhartMarriottMogilnerTheriot2008}; cytoskeletal forces drive flickering in red blood cell membranes~\cite{BrochardLennon1975,GovZilmanSafran2003,FournierLacosteRaphael2004,GovSafran2005,TurlierFedosovAudolyAuthGovSykesJoannyGompperBetz2016}; and actomyosin molecular motors remodel condensates~\cite{Schwayer2019_ZO1,KimKalappurakkalMayorRosen2019,WiegandHyman2020}. Together, these examples illustrate that biological interfaces are not passive materials, but dynamic surfaces whose form and fluctuations are constantly reshaped by active processes.

To uncover the physics of these active interfaces, researchers have turned to simplified experimental model systems that strip away cellular complexity~\cite{adkins_dynamics_2022,tayar_controlling_2023,zhao_asymmetric_2024,takatori_active_2020,vutukuri_active_2020,keber_topology_2014,sciortino_active_2025,park_response_2022,nagard_encapsulated_2022,xie_activity_2022,kokot_spontaneous_2022,ZhaoQiDeng2025,Liu2025Nonequilibrium}. Recent experiments using microtubule–kinesin active nematics demonstrated that activity can drive flat liquid–liquid interfaces far from equilibrium, leading to violation of equipartition~\cite{adkins_dynamics_2022}, propagating interfacial modes at vanishing Reynolds number~\cite{adkins_dynamics_2022}, symmetry-broken shapes~\cite{zhao_asymmetric_2024}, and suppressed phase separation~\cite{tayar_controlling_2023}. Researchers have also studied active vesicles and droplets that mimic cells. In vesicles, internal active stresses produce large, non-equilibrium deformations of elastic membranes~\cite{takatori_active_2020,vutukuri_active_2020,keber_topology_2014,sciortino_active_2025,park_response_2022,nagard_encapsulated_2022}. {In contrast, purely fluid droplets—whose interfaces are governed by surface tension—are predicted by theory and simulations to change shape, move, or even divide when driven by activity~\cite{ruske_morphology_2021,tjhung_spontaneous_2012,tjhung_contractile_2017,giomi_spontaneous_2014,blow_biphasic_2014,Nejad2023,young_many_2021,kawakami2025migration}, and several recent experiments have begun to probe such internally driven liquid droplets~\cite{xie_activity_2022,kokot_spontaneous_2022,ZhaoQiDeng2025,Liu2025Nonequilibrium}.
However, it remains experimentally unclear how a closed, three-dimensional liquid–liquid interface responds when driven internally by active turbulence.}

\begin{figure*}[htp]
\centering
\includegraphics[width=1\textwidth]{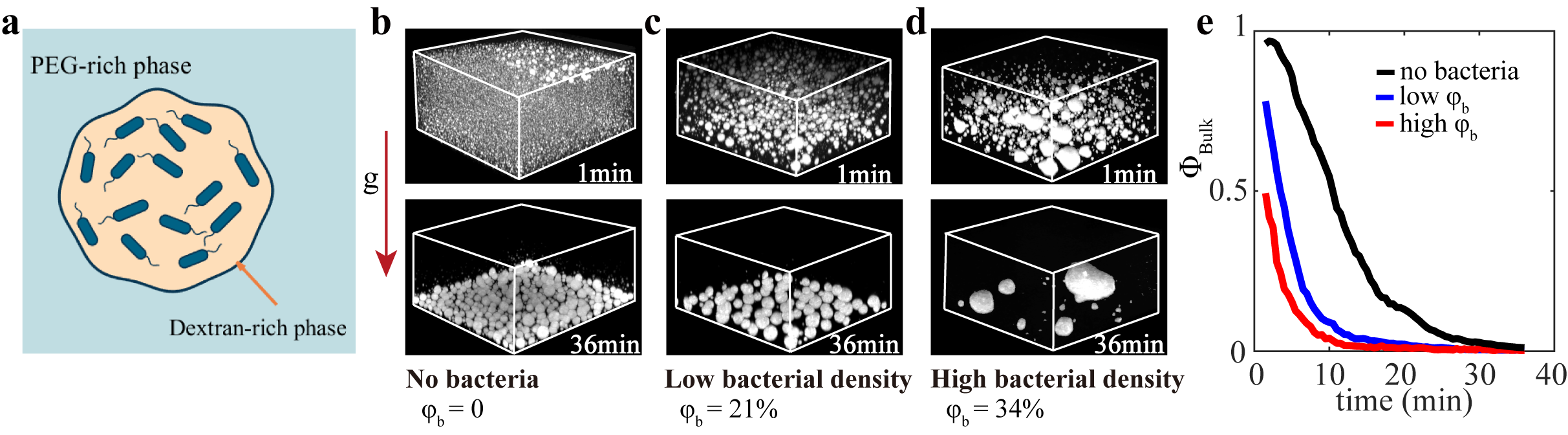}
\caption{ Phase separation and sedimentation dynamics. a, Schematic illustration of an active bacterial droplet. b--d, Confocal images at the initial ($T=1~\mathrm{min}$, top) and final ($T=36~\mathrm{min}$, bottom) stages of coarsening and sedimentation for b, the control condition without bacteria; c, the low-bacterial-density condition ($\varphi_\mathrm{b}=21\pm4\%$); and d, the high-bacterial-density condition ($\varphi_\mathrm{b}=34\pm3\%$). Field of view: $330~\mu\mathrm{m} \times 330~\mu\mathrm{m} \times 200~\mu\mathrm{m}$. Fluorescence highlights the dextran-rich phase. e, Time evolution of the volume fraction of droplets in the bulk, excluding droplets touching the substrate. Measurements of liquid densities, viscosities, contact angles and surface tension are provided in the Supplementary Information. Source data are provided as a Source Data file.}
\label{fig:1}
\end{figure*}

In this work, we introduce an experimental platform of three-dimensional phase-separated liquid droplets encapsulating dense suspensions of motile bacteria. {This system provides a minimal setting in which a closed liquid–liquid interface is driven internally by bacterial turbulence. Combining experiments, simulations, and theory, we identify two distinct dynamical regimes. At lower bacterial density, the interface exhibits propagating capillary-like modes in the Stokesian regime, which we attribute to activity-induced reduction of viscous damping by internal bacterial stresses. At higher bacterial density, bacterial motion drives the interface deep into the nonlinear regime, producing multiscale deformations ranging from micron-scale filaments to protrusions tens of microns in width, including elongated shapes that persist beyond the classical Plateau–Rayleigh instability threshold, while also accelerating droplet motion and coalescence. Together, these results establish dense bacteria-laden liquid droplets as a versatile model system for studying how internally generated active stresses reshape fully fluid interfaces in three dimensions.} More broadly, our results illustrate a wide spectrum of activity-driven interfacial dynamics in living droplets and suggest new routes toward engineering synthetic cells whose mechanics are powered by internal motility~\cite{crowe_liquidliquid_2018,SalehiReyhani2017}.

\section*{Results}
\subsection*{Bacteria-polymer active mixtures}

Our experimental system is a ternary mixture of polyethylene glycol (PEG), dextran, and motile \textit{E.~coli} bacteria (Fig. 1{a}). Without bacteria, the PEG/dextran mixture undergoes liquid-liquid phase separation, forming dextran-rich droplets with a surface tension of {$\gamma_0=7.5\pm1.7~\mu\mathrm{N/m}$ (\textcolor{blue}{Methods}, Supplementary Figure~1), for which the thermal capillary amplitude scale \(\sqrt{k_BT/\gamma_0}\) is only about \(20\text{-}30~{\rm nm}\), far below the micrometer-scale active deformations resolved here.} As shown in Fig. 1{b} and Supplementary Movie 1, these fluorescently labeled dextran-rich droplets coarsen and sediment over time. When motile bacteria are added to the mixture, they fully partition into the more hydrophilic dextran-rich phase~\cite{yaguchi_micropatterning_2010,Cheon2025MotilityPRL}. 
We use two bacterial volume fractions-defined as the ratio of the bacteria volume to that of dextran-rich droplets-with $\varphi_\text{b} = 21\pm4\%$ in Fig. 1{c} and $\varphi_\text{b} = 34\pm3\%$ in Fig. 1{d} (Supplementary Figure~2).

Bacterial motion actively deforms and displaces the droplets, which dramatically accelerates sedimentation: all bacteria-laden droplets reach the chamber bottom within 10 minutes (Fig. 1e). Once settled, these droplets-with radii $R$ ranging from tens to hundreds micrometers-serve as confined environments in which we directly measure bacterial motion and characterize the interfacial dynamics that emerge at the liquid-liquid phase boundary.
{ Under both volume fractions, the encapsulated bacteria exhibit turbulence-like collective motion~\cite{Peng2021BacterialTurbulence,Sokolov2007CollectiveDynamics,dunkel_fluid_2013}. Increasing \(\varphi_\text{b}\) increases the collective active forcing generated per unit droplet volume, and therefore corresponds to a higher collective activity of the bacterial suspension (Supplementary Figure~4). As we show below, these two density conditions lead to qualitatively distinct regimes of interfacial dynamics.}

\subsection*{Activity-driven interfacial fluctuations}

At low bacterial density ($\varphi_\text{b} = 21\pm4\%$), dextran-rich droplets remain approximately spherical (Fig. 2a, Supplementary Figure~3a and Supplementary Movie 2). We image bacterial motion on the equatorial plane of the droplets. As shown in Fig. 2a–Fig. 2b and Supplementary Movie 3, collective bacterial flows emerge inside two representative droplets. From the measured velocity fields, we compute the velocity auto-correlation function and extract a characteristic correlation length $\ell_\mathrm{v}$ (Supplementary Figure~17a). The value of $\ell_\mathrm{v}$ is on the order of tens of micrometers and grows sublinearly with droplet size $R$ (Supplementary Figure~17b), consistent with recent experimental observations of confined bacterial turbulence~\cite{wei_scaling_2024,PerezEstay2026ScaleFree}.

\begin{figure*}[htp]
\centering
\includegraphics[width=1\textwidth]{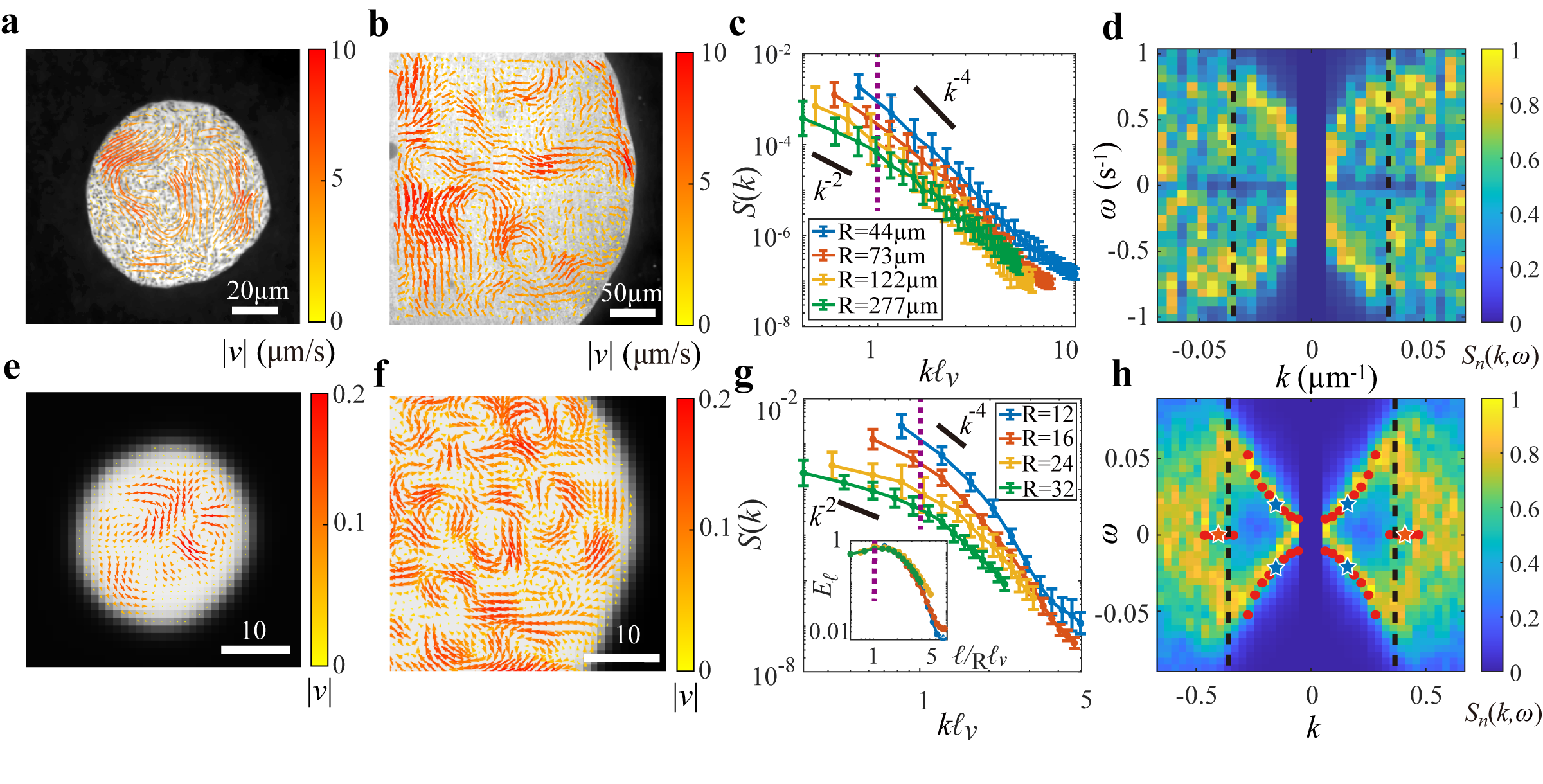}
\caption{Shape dynamics of active droplets measured in experiments and simulations. a,b, Experimental bacterial flow fields measured on the equatorial plane in two droplets with different radii: a, $R=39.6~\mu\mathrm{m}$ and b, $R=273~\mu\mathrm{m}$. c, Fluctuation spectra, $S(k)$, measured in four representative droplets with different radii. Each curve corresponds to one individual droplet. For each droplet, one long acquisition is divided into 30 independent temporal segments, and the plotted spectrum and error bars represent the mean and standard deviation over the segment-resolved spectra, respectively. Error bars are displayed symmetrically in logarithmic space. The horizontal axis is normalized by the corresponding flow correlation length, $\ell_\mathrm{v}$. The dashed line marks $k\ell_\mathrm{v}=1$. d, Normalized dynamic structure factor, $S_\mathrm{n}(k,\omega)$, measured in a droplet with $R=226~\mu\mathrm{m}$. The dashed line marks the cutoff wavevector, $k_c$. e,f, Numerical bacterial flow fields measured on the equatorial plane in two simulated droplets with different radii, reported in simulation units: e, $R=12$ and f, $R=32$. g, Fluctuation spectra, $S(k)$, obtained from simulations of four droplets. Inset: mean interfacial energy $E_\ell$ as a function of $\ell R^{-1}\ell_\mathrm{v}$, averaged over $m$. The dashed lines mark $k\ell_\mathrm{v}=1$. h, Normalized dynamic structure factor, $S_\mathrm{n}(k,\omega)$, for a simulated droplet with $R=32$. The dashed line marks the cutoff wavevector, $k_c$. Red dots denote characteristic frequencies extracted from simulations with specific initial conditions (see Fig.~3), and stars correspond to the two curves with $\ell=5,m=0$ (blue) and $\ell=13,m=0$ (orange) in Fig.~3b. The zero-frequency points correspond to non-oscillatory damping conditions. All experimental data are obtained in droplets with $\varphi_\mathrm{b}=21\%$. All length and velocity scales in the simulations are reported in dimensionless simulation units. Their conversion to physical units is described in Supplementary Information, Sec.~S2C. Source data are provided as a Source Data file.}
\label{fig:2}
\end{figure*}

Bacterial motion induces pronounced droplet interface fluctuations (Fig. 2a–Fig. 2b and Supplementary Movie 3), which are negligible in similarly sized passive droplets at thermal equilibrium.
 To quantify the interface fluctuations, we parameterize the equatorial contour at time $t$ as $r(\varphi,t)$, where $\varphi$ is the azimuthal coordinate (Supplementary Figure~6). We then compute the temporally averaged fluctuation spectrum $S(k) = \langle|\hat{u}_{k}|^{2}\rangle$, with $\hat{u}_{k}$ the spatial Fourier transform of the radial deformation $r(\varphi,t)$~\cite{pecreaux_refined_2004}. As shown in Fig. 2c for four representative droplets, the spectra exhibit a clear scaling transition near the characteristic wavevector of $1/\ell_\mathrm{v}$: from a $k^{-2}$ dependence at small $k$ to a $k^{-4}$ dependence at large $k$. This scale-dependent behavior underscores how nonequilibrium driving fundamentally redistributes interfacial energy among modes~\cite{takatori_active_2020,vutukuri_active_2020,park_response_2022,adkins_dynamics_2022,kokot_spontaneous_2022}, in contrast to the equilibrium-like equipartition observed in passive droplets or vesicles~\cite{helfrich_elastic_1973,stottrup_determination_2007,pecreaux_refined_2004,honerkamp-smith_line_2008,shimobayashi_critical_2023}. Temporal behavior of the fluctuations can be characterized by the normalized dynamic structure factor \( S_\mathrm{n}(k,\omega) \) (\textcolor{blue}{Methods}), which exhibits peaks at nonzero frequencies for small wavenumbers (Fig.~\ref{fig:2}d). This indicates the presence of propagating wave modes, consistent with the kymographs in Supplementary Figure~9a. As \( k \) increases,  \( S_\mathrm{n}(k,\omega) \)  no longer displays distinct wave peaks. This transition occurs at a \( k_c \), marked by the dashed line in Fig.~\ref{fig:2}d, which approximately corresponds to the scale of the velocity correlation length $\ell_\mathrm{v}$.

{To further investigate the spatiotemporal dynamics of the active droplets, we simulate a minimal coarse-grained continuum model of an active–passive fluid mixture that reproduces the key qualitative features of the experiments.} Our model couples the Cahn-Hilliard equation for phase separation with the Toner-Tu-Swift-Hohenberg (TTSH) equation for bacterial motion ({Methods})~\cite{yue2004diffuse,dunkel_fluid_2013,dunkel_minimal_2013,heidenreich_hydrodynamic_2016,reinken_derivation_2018}. Parameter choices and simulation details are given in Sec. S2 in the SI. Our simulations reproduce the experimental observations of collective bacterial flows and interface fluctuations (Fig. 2{e}, Fig. 2{f}). To analyze the interface dynamics, we represent the droplet interface as $r(\theta,\varphi)$, where $\theta$ and $\varphi$ are the polar and azimuthal angles, respectively. We then decompose it into spherical harmonics, $Y_{\ell,m}(\theta,\varphi)$, specified by two indices, $(\ell,m)$. From this decomposition, we compute the 3D fluctuation spectra and the mean interfacial energy spectrum $E_{\ell}$ (Fig. 2{g} inset, Supplementary Figure~7a-b). When $E_{\ell}$ is plotted against $\ell/R$ for  droplets of different radii $R$, the data collapse onto a single curve with a distinct peak at $1/\ell_\mathrm{v}$ ( Fig. 2{g} inset). This peak reflects the scale at which active flows most strongly energize the interface, coinciding with the velocity correlation length $\ell_\mathrm{v}$ known from studies of active turbulence~\cite{Wensink2012MesoScale,Peng2021BacterialTurbulence}. Projecting the spherical harmonic spectra onto the equatorial plane yields the 2D fluctuation spectra $S(k)$ shown in Fig. 2{g}. These spectra are consistent with the experimental results in Fig. 2c and exhibit the same transition in scaling behavior at $1/\ell_\mathrm{v}$. 
Thus, experiments and simulations together paint a consistent picture: active flows inject energy preferentially at the scale $\ell_\mathrm{v}$, driving large-scale interfacial motion ($k^{-2}$), while small-scale fluctuations decay more steeply ($k^{-4}$). {This activity-dominated crossover is distinct from passive equilibrium crossovers: unlike vesicle membranes~\cite{helfrich_elastic_1973,pecreaux_refined_2004,takatori_active_2020,vutukuri_active_2020}, the liquid-liquid interface has surface tension but no bending rigidity, and thermal fluctuations are negligible compared with the activity-driven interfacial motion observed here.}

\begin{figure}[tp]
\centering
\includegraphics[width=0.95\columnwidth]{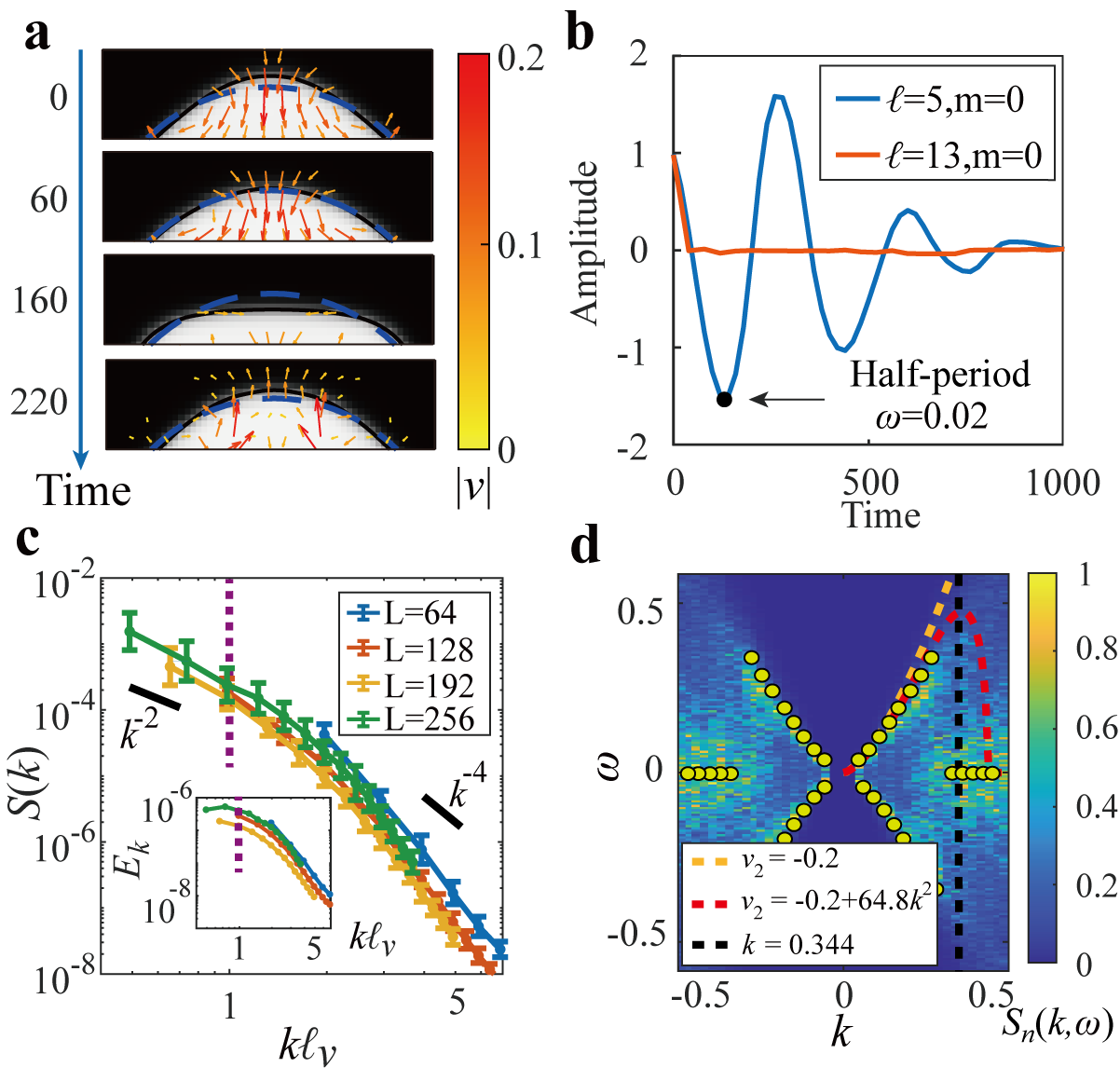}
\caption{Wave dynamics from simulations of a spherical interface and a flat interface. a, Interface evolution and associated flow field resulting from an initial perturbation with the spherical harmonic mode $(\ell=5,m=0)$ on a spherical interface. The blue dashed line indicates the unperturbed shape. One fifth of the droplet is shown for clarity. b, Temporal evolution of the amplitudes of the $(\ell=5,m=0)$ and $(\ell=13,m=0)$ modes. The black dot denotes the corresponding frequency appearing in Fig.~2h. c, Fluctuation spectra, $S(k)$, obtained from two-dimensional simulations of four flat interfaces. Inset: mean interfacial energy $E_\mathrm{k}$ as a function of $k\ell_\mathrm{v}$. The dashed lines mark $k\ell_\mathrm{v}=1$. The expression of $E_\mathrm{k}$ is provided in Supplementary Information, Sec.~S3. d, Normalized dynamic structure factor, $S_\mathrm{n}(k,\omega)$, measured from a flat interface. The orange and red dashed lines represent the analytical results in Eq.~\eqref{eq:oscillator} without and with the inclusion of higher-order viscosity, respectively. The black dashed line marks the threshold beyond which propagating modes disappear. The yellow dots denote characteristic frequencies extracted from simulations with specific initial conditions. Details of the initial-perturbation analysis are provided in Supplementary Figure~20d-f. All simulation parameters are given in Supplementary Information, Sec.~S2. Source data are provided as a Source Data file.}
\label{fig:3}
\end{figure}

Our simulations also capture the temporal behavior seen in experiments. Kymographs of equatorial fluctuations \(r(\varphi,t)\) (Supplementary Figure~10a) and the dynamic structure factor (Fig. 2{h}) both closely match their experimental counterparts, confirming the presence of propagating wave modes at small \(k\). To understand how such waves can arise at low Reynolds number, we impose a small-amplitude deformation on the interface corresponding to a spherical harmonic mode with \( (\ell\neq0, m=0) \)(Supplementary Figure~18 and \textcolor{blue}{Methods}). These axisymmetric modes can be projected onto the meridional plane, yielding an effective deformation wave vector \(k = \ell / R\). Figure \ref{fig:3}{a} shows how the interface and the surrounding fluid respond after an initial \( (\ell=5, m=0) \) perturbation. The mode amplitude oscillates over time (Fig.~\ref{fig:3}{b}), indicating a wave-like response. The corresponding flow fields in Fig.~\ref{fig:3}{a} reveal an effective inertial effect that sustains the oscillations: in a passive low-Re fluid, capillary-driven flow stops once the interface relaxes to equilibrium; here, active bacterial motion sustains the flow even at equilibrium, as if the fluid possessed inertia. This persistent flow pushes the interface past its equilibrium shape, driving oscillatory dynamics. Initial perturbations with large $\ell$ decay rapidly without oscillation (Fig.~\ref{fig:3}{b}), consistent with the diffusive (zero-frequency) modes seen at high $k$ in Fig. \ref{fig:2}h. By systematically varying the spherical harmonic degree $\ell$ of the initial perturbation, we measure the oscillation frequency for each mode (Fig.~\ref{fig:3}{b} and Supplementary Figure~20c). These frequencies—plotted as red dots in Fig. 2h—align closely with the peaks in $S_{n}(k,\omega)$ computed from the full simulation, confirming that the propagating waves are sustained by persistent active flows. This dynamic structure, including wave propagation at low $k$ and the crossover to diffusive behavior, remains robust across a wide range of parameters (Supplementary Figure~24).

To test the universality of the wave mechanism and to enable analytical progress, we use our simulations to investigate interfacial fluctuations on a flat interface separating passive and active fluids in two dimensions (Supplementary Figure~20). The fluctuation spectra of the 2D interface show a clear transition in scaling at $1/\ell_\mathrm{v}$ (Fig. 3{c}), marking the scale at which active flows most strongly energize the interface (Fig. 3{c} inset). The normalized dynamic structure factor $S_{n}(k,\omega)$ of the flat interface resembles its spherical counterpart, showing clear evidence of propagating modes at $k<k_c$ (Fig. 3{d}). Oscillation frequencies extracted from perturbed flat interfaces match those obtained from the turbulent steady state (Fig. 3{d} and Supplementary Figure~20d-f). Together, these results confirm that the wave mechanism and the role of active flows are not specific to spherical geometry—they also govern planar interfaces in 2D.

The 2D planar geometry greatly simplifies the analytical treatment. To linear order, the interface height $h(x,t)$ evolves according to the normal  fluid velocity at the interface: $\partial_{t}h(x,t)=v_{y}(x,y=0,t)$. The flow fields in each phase are described by the linearized TTSH equation
\begin{equation}
\partial_{t}\mathbf{v}^{i}=-\nabla P^{i}-\alpha_{i}\mathbf{v}^{i}+\nu_{i}\nabla^{2}\mathbf{v}^{i}\,,
\label{eq:nseq}
\end{equation}
where $i=1,2$ denote passive and active fluids, $\alpha_{i}$ is the effective damping coefficient, and $\nu_{i}$ is the effective viscosity. For analytical tractability, we neglect the fourth-order viscosity term. Solving Eq.~\eqref{eq:nseq} with proper boundary conditions at the interface yields the dispersion relation of the interfacial waves, which agrees quantitatively with simulations for $k<k_c$ (Supplementary Figure~28a, see \textcolor{blue}{Methods} and Sec.~S4 for details). In the limit of small $\alpha_{i}$ and $|\nu_{i}|k^{2}\gg\omega$, both justified in Sec.~S4F, the dispersion relation reduces to a simple form
\begin{equation}
\label{eq:oscillator}
    (i\tilde{\omega})^2+\frac{1}{\tau_\mathrm{v}(k)}i\tilde{\omega}+\frac{1}{\tau_{0}(k)^2} = 0,
\end{equation}
which is analogous to a damped harmonic oscillator. The interface dynamics are governed by two characteristic time scales: the viscous damping time $\tau_\mathrm{v}(k)=3/[2(\nu_{1}+\nu_{2})k^{2}]$ and the capillary-driven oscillation period $\tau_{0}(k)=\sqrt{3/(\gamma_{0}k^{3})}$. As in a damped oscillator, oscillatory modes arise in the underdamped region, $\tau_\mathrm{v}>\tau_{0}/2$. In passive low-Reynolds-number droplets, viscous damping dominates so that $\tau_\mathrm{v}\ll\tau_{0}$ and interfacial modes are purely diffusive. In contrast, droplets encapsulating bacteria acquire a negative effective viscosity $\nu_{2}$ in the TTSH theory. This negative $\nu_{2}$ can offset viscous damping exerted by the surrounding passive phase on the interface, when $\nu_{1}+\nu_{2}$ is sufficiently close to zero, effectively underdamping the interface and generating inertia-like dynamics. The underdamped condition, $\tau_\mathrm{v}>\tau_{0}/2$, also defines a critical wavenumber $k_c'=3\gamma_0/(\nu_1+\nu_2)^2$ beyond which interfacial modes become diffusive (see Sec. S4E). This theoretical cutoff, however, overestimates the value of $k_c$ observed in simulations, due to our neglect of higher-order viscosity and nonlinear effects. We incorporate these missing effects phenomenologically by introducing a scale-dependent active viscosity \( \nu_2'(k) \equiv \nu_2(1 - 4\ell_\mathrm{v}^2 k^2) \), which reduces active driving at wavelengths smaller than the typical vortex size \( 2\ell_\mathrm{v} \). With this adjustment, the predicted dispersion relation and cutoff wavevector  \( k_c' \) both agree with the simulation results (red curve in Fig. 3{d}, Supplementary Figure~28b).

\begin{figure}[tp]
\centering
\includegraphics[width=0.48\textwidth]{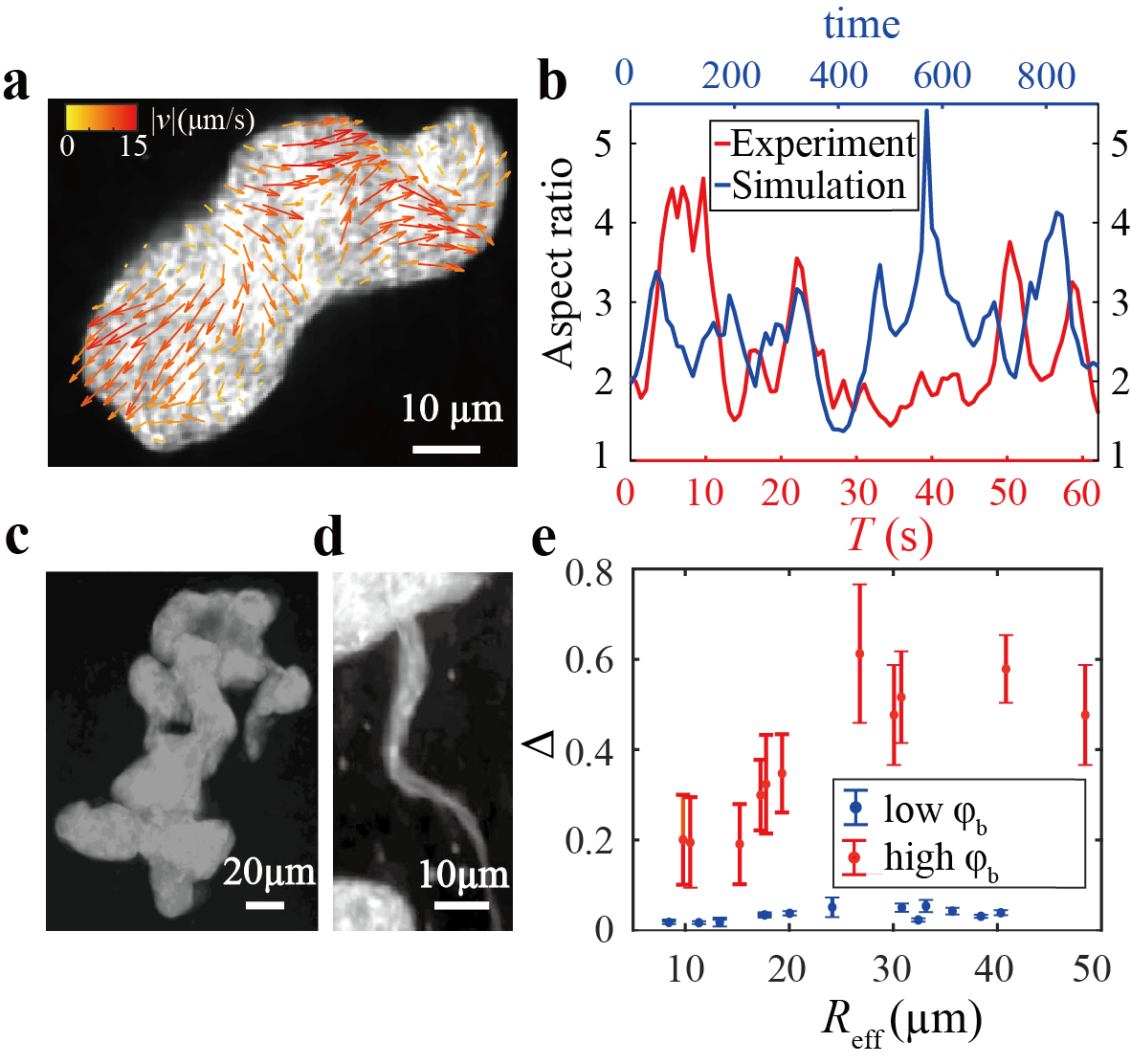}
\caption{Shape and dynamics of strongly deformed droplets. a, Bacterial flow field in an elongating droplet. b, Temporal evolution of the droplet aspect ratio in experiment (red) and simulation (blue). c, Representative image of a droplet exhibiting a complex, highly non-spherical shape. d, Filament-like protrusions extending from the droplet interface. e, Normalized surface area as a function of droplet size, $R_\mathrm{eff}$, under low-bacterial-density (blue) and high-bacterial-density (red) conditions. $R_\mathrm{eff}$ is defined as $R_\mathrm{eff}=(3V/4\pi)^{1/3}$, where $V$ is the droplet volume. Error bars indicate the standard deviation over a $1~\mathrm{min}$ measurement window. All experimental data correspond to droplets with $\varphi_\mathrm{b}=34\%$, except for the low-bacterial-density data in e. The simulation curve in b corresponds to the snapshots and parameters in Supplementary Figure~30. Source data are provided as a Source Data file.}
\label{fig:4}
\end{figure}

\subsection*{Large droplet deformations}

When the bacterial concentration is increased to $34\%$, strong internal flows generate persistent protrusions that push the interface outward. The resulting deformations in turn steer the active flows, creating a positive feedback loop that amplifies droplet elongation. This coupling produces highly anisotropic droplet shapes with characteristic widths on the order of the flow correlation length, $\ell_\mathrm{v}$, as shown in Fig. 4{a}, Supplementary Figure~3b, Supplementary Figure~29–32, Supplementary Figure~37, and Supplementary Movie~5–6. The aspect ratio evolution in both experiment and simulation shows stretching beyond the Rayleigh–Plateau threshold (aspect ratio $\pi$) for several seconds, well above the passive breakup timescale $6\eta R/\gamma_0 \sim 0.1~ \mathrm{s}$ (Fig. 4{b} and Supplementary Movie 6) ~\cite{rayleigh1892xvi,plateau1873statique,pekker_plateaux2013rayleigh_2018}. {These observations suggest that internal active forcing can transiently maintain highly elongated interfacial configurations beyond the passive capillary-breakup expectation.}

In large droplets, strong internal flows generate multiple protrusions, producing a broad spectrum of deformation modes and complex, time-evolving morphologies (Fig. 4{c} and Supplementary Movie 7). To measure shape deviation from the spherical equilibrium favored by surface tension, we define the normalized excess surface area, $\Delta \equiv S/S_{0} - 1$, where $S$ is the droplet surface area and $S_{0}$ is that of a volume-matched sphere. Under low-bacterial-density conditions, $\Delta$ remains near zero (Fig. 4{e}), reflecting near-spherical shapes. At high bacterial density, however, large droplets exhibit substantial excess areas, with $\Delta$ reaching $\sim 0.6$, corresponding to a $60\%$ increase in surface area relative to a sphere (Supplementary Figure~34).

On the surfaces of highly active droplets, we also observe thin protruded filaments with widths comparable to that of a single bacterium (Fig. 4{d}, Supplementary Figure~3d, Supplementary Figure~35–37, and Supplementary Movie 8). Within these filaments, bacteria align nematically along the filament axis (Supplementary Figure~36), indicating localized ordering that emerges from the particulate character of the suspension. These filaments can sometimes detach from the droplet surface. These features illustrate the multiscale nature of bacterial droplets, where microscopic alignment and mesoscale flow structures coexist and cooperatively shape the interface~\cite{takatori_active_2020,prasad_alcanivorax_2023}.
{Together, the interfacial wave modes, protrusions, and bacteria-scale filaments show that bacterial droplets do not follow a single interfacial response mechanism. Instead, they exhibit activity-dependent regimes ranging from near-linear propagating waves to nonlinear large-scale deformation and localized bacterial alignment.}

\subsection*{Contact line, droplet motion and coarsening}

Passive droplets that sediment to the chamber bottom form stable contact lines with the substrate, establishing a contact angle of about  $150^{\circ}$ (Supplementary Figure~11a). Active droplets behave very differently. Under low bacterial volume fraction, droplets retain a similar contact angle (Supplementary Figure~11b), but bacterial flows deform the contact line (Fig. 5a, Fig. 5b, Supplementary Figure~12), producing small yet persistent lateral displacements on the substrate (Fig. 5c). Under higher bacterial volume fraction, strong interfacial fluctuations eliminate any well-defined static contact angle (Supplementary Figure~11c) and generate highly irregular contact lines, intermittent detachment, and substantial lateral motion (Fig. 5d–Fig. 5f, Supplementary Figure~13). This active migration near the substrate strongly influences coarsening dynamics (Supplementary Movie 9).

To quantify this substrate-mediated coarsening, we measure the correlation length \(\xi\) of fluorescence intensity fluctuations within a \(40~\mu\mathrm{m}\) layer above the substrate, which reflects the typical droplet size near the surface. As shown in Fig. 5g, \(\xi\) initially increases as droplets sediment, encounter one another, and coalesce near the bottom surface. In passive and low-bacterial-volume-fraction samples, however, coarsening slows down and eventually saturates once the droplets become constrained by their contact lines (Fig. 1b-c, Fig. 5h). In contrast, at high bacterial volume fraction, bacterial flows strongly deform the interface and contact line, allowing droplets to move, partially detach, and continue encountering neighboring droplets. This sustained substrate-level mobility promotes further fusion, leading to a steadily increasing \(\xi\) and a final state dominated by a few large droplets (Fig. 1d, Fig. 5h).

\begin{figure}[tp]
\centering
\includegraphics[width=0.49\textwidth]{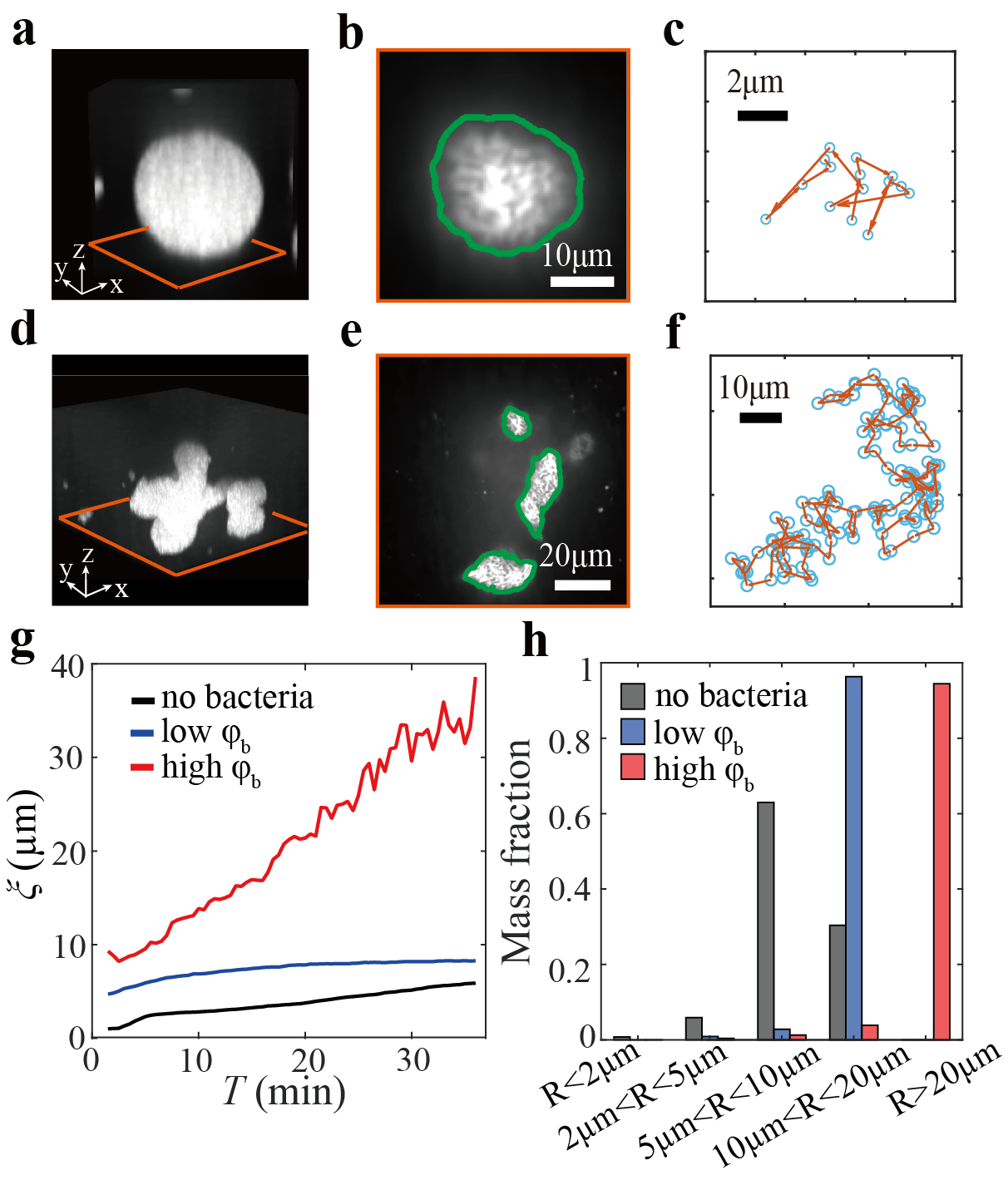}
\caption{Droplet motion and coarsening near the substrate. a, Side view of a low-bacterial-density droplet with $\varphi_\mathrm{b}=21\%$. The red contour marks the substrate position. b, Contact line corresponding to the droplet in a. c, Trajectory of the droplet center of mass with a time interval of $4.7~\mathrm{s}$. d, Side view of a high-bacterial-density droplet with $\varphi_\mathrm{b}=34\%$, showing enhanced interfacial deformation. The red contour marks the substrate position. e, Contact line corresponding to the droplet in d. f, Trajectory of the droplet center of mass with a time interval of $0.87~\mathrm{s}$. g, Time evolution of the characteristic length of the dextran phase, $\xi(t)$, within a $40~\mu\mathrm{m}$ region above the substrate for the zero-, low-, and high-bacterial-density conditions. h, Droplet size distributions at $T=36~\mathrm{min}$ for the zero-, low-, and high-bacterial-density conditions. Source data are provided as a Source Data file.}
\label{fig:5}
\end{figure}

\section*{Discussion}

Activity is increasingly recognized as a powerful driver of interfacial and membrane physics, endowing soft boundaries with distinct nonequilibrium dynamics~\cite{vutukuri_active_2020,kokot_spontaneous_2022,takatori_active_2020,park_response_2022,sciortino_active_2025,adkins_dynamics_2022,gulati_traveling_waves_active_liquid_crystals_2024,nagard_encapsulated_2022,prasad_alcanivorax_2023,Cates2025,tayar_controlling_2023,caballero_activity-suppressed_2022,bhattacharyya_phase_2023,blow_biphasic_2014,ruske_morphology_2021,li_shape_transformations_2019,CarenzaGiomi2022_DefectMediatedMorphodynamics,fausti_capillary_2021,zhao_asymmetric_2024,keber_topology_2014,xie_activity_2022,ZhaoQiDeng2025,tjhung_contractile_2017,tjhung_spontaneous_2012,giomi_spontaneous_2014,Nejad2023,young_many_2021,kawakami2025migration,Liu2025Nonequilibrium,bar-ziv_pearling_1999,BarZiv1994Instability,Peterson2021VesicleST}. Here,  we establish an experimental platform for bacteria-powered, phase-separated droplets and show how internal activity reshapes their dynamics, shape, and  coarsening behavior. 
In the low bacterial volume fraction regime, the droplet interface shows scale-selective fluctuations. Energy is injected at the length scale of bacterial turbulence, producing a distinctive spectral crossover in both experiments and simulations (Fig. 2c, Fig. 2{g}, Supplementary Figure~7 and 8). These scale-dependent fluctuations arise from collective bacterial flows~\cite{adkins_dynamics_2022}, in contrast to the broadband, uncorrelated agitation generated by isolated active units~\cite{vutukuri_active_2020,takatori_active_2020,sciortino_active_2025}. 
Moreover, the coupling between active and passive fluids restores effective inertia even at vanishing Reynolds number: active stresses counteract viscous damping on the interface and enables underdamped, capillary-like waves on a liquid–liquid boundary (Fig. 2{d}, Fig. 2{h}, Fig. 3{a}, Fig. 3{d}). 
{
Interfacial waves in active nematics are typically attributed to nonreciprocal coupling between the nematic director and the interface ~\cite{adkins_dynamics_2022,gulati_traveling_waves_active_liquid_crystals_2024}. Traveling membrane deformations in active vesicles, on the other hand, occur on elastic boundaries, where membrane bending contributes to the restoring force ~\cite{sciortino_active_2025}. In contrast, our results suggest a continuum-level hydrodynamic mechanism in which activity-induced viscosity reduction in the bacterial phase compensates viscous damping from the passive phase.
 This suggests that internal motility can convert a soft boundary from purely dissipative to oscillatory—a principle useful for designing wave-supporting interfaces in living and synthetic systems (SI Sec. S4G).}

At high bacterial volume fraction, internal stresses generate large, persistent droplet deformations (Fig. 4{a}, Fig. 4{b}). Such extreme shapes are typically unstable in liquid–liquid interfaces, where Rayleigh–Plateau instabilities promote breakup, unlike membrane-bound systems where bending rigidity prevents rupture~\cite{takatori_active_2020,nagard_encapsulated_2022,prasad_alcanivorax_2023,bar-ziv_pearling_1999,BarZiv1994Instability}. 
{These observations suggest that internal activity may help maintain strongly deformed, connected interfacial configurations that have no direct passive counterpart.}
Activity also accelerates coarsening. 
While previous studies have shown that activity can regulate domain growth and induce microphase separation~\cite{Cates2025,tayar_controlling_2023,caballero_activity-suppressed_2022,bhattacharyya_phase_2023,blow_biphasic_2014,fausti_capillary_2021}, 
{our system further involves gravitational sedimentation and a bottom substrate. 
In the bulk, internal active flows deform droplets and enhance transport, increasing droplet encounters and coalescence. As droplets grow, faster sedimentation further promotes collisions (SI Sec. S3D). 
On the substrate, passive droplets tend to be pinned by stable contact lines, whereas high bacterial concentration droplets remain internally driven, allowing contact-line deformation, detachment from the substrate and lateral motion.
Together, these effects provide an activity-assisted pathway for accelerated coarsening under gravity and confinement.}
These effects establish motility as a powerful control parameter for both morphology and phase-separation kinetics in active emulsions.

Overall, our results show that internal active stresses can fundamentally reprogram interface mechanics in phase-separated droplets. Active droplets thus behave as self-powered mechanical compartments whose interfaces resist rupture, sustain large deformations, and exhibit motility—traits typically associated with living cells. These behaviors emerge from nonequilibrium energy injection and fluid–interface coupling, illustrating how liquid boundaries can acquire mechanical function from within. Our findings provide a foundation for designing programmable active droplets and mechanically functional artificial cells.



\section*{Methods}

\subsection*{Bacterial cultivation}
We use \textit{E. coli} BW25113 as the active component. Cells are cultured overnight at 37.0~°C in Terrific Broth (TB), diluted 1:100 in fresh TB medium, and grown at 30.0~°C for 5 hours. After centrifugation and washing with motility buffer, the cells are resuspended and prepared for further experiments. 

\subsection*{Phase-separating solution}
The phase-separating mixture is prepared by dissolving dextran (DEX) and polyethylene glycol (PEG) at equal initial mass concentrations in motility buffer.
Fluorescein-labeled dextran (D1844, 40,000~MW) is added at a final concentration below 0.1\% (w/w) to enable fluorescent visualization of the dextran-rich phase. The interfacial tension of the resulting liquid–liquid interface is measured to be $7.5\pm1.7~\mu$N/m. 

\subsection*{Sample preparation}

To set the desired bacterial volume fraction, we mix 200~$\mu$L of PEG-rich phase, 10~$\mu$L of DEX-rich phase, and either 5~$\mu$L (low bacterial density, $\varphi_\mathrm{b} = 21\pm4\%$) or 8~$\mu$L  (high bacterial density, $\varphi_\mathrm{b} = 34\pm3\%$) of dense bacterial suspension. The bacterial volume fraction is estimated by scanning the droplets under a 240× SORA mode on a confocal microscope and manually counting bacteria. Samples are sealed in circular chambers (radius: $2.5~\text{mm}$) assembled from double-sided adhesive tape sandwiched between two glass slides. The chamber height is set to $200~\mu$m for small droplets and coarsening measurements, and to 
$600~\mu$m for samples containing droplets larger than $100~\mu$m in radius, preventing contact with the upper slide.

\subsection*{Imaging}

Imaging is performed using a Nikon spinning disk confocal microscope (CSU-W1). Flow-field measurements and contour extraction for droplets with radii below
$100~\mu$m use a 60× water-immersion objective. Droplets larger than $100~\mu$m are imaged using a 20× objective. 

\subsection*{Data analysis}

The acquired images are preprocessed using the denoise.ai module in NIS-Element Viewer. Droplet contours are identified in MATLAB via binarization. Morphological properties of 3D droplets—including surface area, volume, and aspect ratio—are extracted using MATLAB’s \texttt{regionprops3}.

The 2D deformation field $r(\varphi,t)$ is defined as the radial distance from each contour pixel to the center of mass of the 2D cross-section. Fluctuation spectra are computed using discrete Fourier transforms. The dynamic structure factor (DSF) is calculated by performing a spatial-temporal Fourier transform after interpolating the boundary points $(r_{i},\varphi_{i})$ at each time $t$ onto a uniform grid. Because the fluctuation spectrum decays with increasing $k$, the DSF is dominated by the largest spatial scales (Supplementary Figure~9b and 10b). To consistently compare modes with different amplitudes, we normalize each DSF curve at fixed 
$k$ by its maximum value. Specifically, at $k = k_i$, the normalized DSF $S_\mathrm{n}(k_i,\omega)$ is defined as
\begin{equation}
S_\mathrm{n}(k_i,\omega) = \frac{S(k_i,\omega)}{\max_{\omega} S(k_i,\omega)}.
\end{equation}

\subsection*{Numerical simulation}
{To further investigate the spatiotemporal dynamics of the active droplets, we simulate a minimal coarse-grained continuum model that captures the coupling between internal bacterial turbulence and a deformable liquid-liquid interface. In this minimal model, the droplet is treated as a closed liquid-liquid interface, while substrate-induced hydrodynamic and contact-line effects are not explicitly included.}
The model comprises a phase field $\phi(\mathbf{x},t)$ and an incompressible flow field $\mathbf{v}$ with $\nabla \cdot \mathbf{v} = 0$. The phase field $\phi(\mathbf{x},t)$ evolves according to an advective Cahn-Hilliard equation:
\begin{equation}
\left(\partial_t + \mathbf{v} \cdot \nabla \right) \phi = \Gamma_\phi \nabla^2 \mu, \label{eq:phi}
\end{equation}
where $\Gamma_\phi$ sets the relaxation rate. The chemical potential  $\mu \equiv \delta F_\phi / \delta \phi$ is derived from a standard Ginzburg–Landau free energy:
\begin{equation}
F^{\phi} = \int_{\mathbf{r}} \left[ \frac{a_\phi}{4} \phi^2 (1-\phi)^2 + \frac{k_\phi}{2} |\nabla \phi|^2 \right] d\mathbf{r}.
\end{equation}
The free energy drives spontaneous phase separation into $\phi = 1$ (active) and $\phi = 0$ (passive) domains, separated by an interface across which $\phi$ varies smoothly. 

The flow field $\mathbf{v}$ is governed by the Toner-Tu-Swift-Hohenberg (TTSH) equation~\cite{dunkel_minimal_2013,dunkel_fluid_2013,heidenreich_hydrodynamic_2016,reinken_derivation_2018}:
\begin{equation}
\label{eq:flow}
\begin{aligned}
\partial_t \mathbf{v} + \lambda_0 \mathbf{v} \cdot \nabla \mathbf{v}
&= -\nabla p^*
- \alpha\!\left( \phi - \frac{\mathbf{v}^2}{v_0^2} \right)\mathbf{v} \\
&\quad + \nu \nabla^2 \mathbf{v}
- \zeta \nabla^4 \mathbf{v}
+ \mu \nabla \phi.
\end{aligned}
\end{equation}
The coefficients in Eq. \eqref{eq:flow} are set to depend on $\phi$ to disable activity in the passive phase $\phi = 0$. Specifically, we choose the advection coefficient as $\lambda_0 = 1 - S_v \phi$, so that it equals 1 in the passive phase and $1 - S_v$ in the active phase, reflecting active self-advection. The parameter $S_v$ is positive for pusher-type swimmers such as the bacteria in our experiment. The modified pressure $p^* = \left(p - \phi S_v\mathbf{v}^2/d\right)$ includes an active contribution in $d$-dimensional space. $\alpha$ and $v_0$ set the characteristic flow speed, enforcing $v \approx 0$ in the passive phase and $v \approx v_0$ in the active phase. Bacterial activity also enters through an activity-dependent viscosity,
\[
\nu = \left[ \nu_1 (1-\phi) + \nu_2 \phi \right],
\]
where $\nu_1$ and $\nu_2$ are the passive and active viscosity, respectively. A negative $\nu_2$ can destabilize the flow and, together with the stabilizing fourth-order viscosity, it sets a characteristic velocity correlation length $\sqrt{\zeta / |\nu_2|}$. In addition to setting the flow length scale, \(\nu_2\) and \(\zeta\) also define the characteristic active-flow velocity scale \(\sqrt{-\nu_2^3/\zeta}\). 
In the simulations, $v_0$ is chosen to be of the same order as the intrinsic active-flow velocity scale \(\sqrt{-\nu_2^3/\zeta}\), so that the nonlinear speed saturation and the active-flow instability are controlled by comparable velocity scales.
Interfacial tension enters via the term $\mu \nabla \phi$~\cite{gurtin1996two, yue2006phase, kim2012phase}. The parameter values and the specific simulation setting are provided in the SI. {To compare with experiments, the interfacial spectra are computed from equatorial contours, consistent with the experimental fluctuation analysis.}

\subsection*{Theory for a flat interface}
In the analytical model, we treat the horizontal interface as a sharp boundary rather than a diffuse phase field~\cite{adkins_dynamics_2022,Grant1983Fluctuating,Harden1991Hydrodynamic,Flekky1996Fluctuating}. The interface deformation is described by a height field $h(x,t)$, separating a passive fluid (fluid 1, above) from an active fluid (fluid 2, below). Activity is incorporated as a negative effective viscosity in the active layer, consistent with the TTSH framework. The linearized hydrodynamic equations in each fluid are given by
\begin{equation}
\partial_t \mathbf{v}^i=- \nabla P^i- \alpha_i \mathbf{v}^i+\nu_i \nabla^2 \mathbf{v}^i, \quad \nabla \cdot \mathbf{v}^i = 0 , \label{eq:nseq2}
\end{equation}
for $i=1,2$. Considering small deformations, the kinematic boundary condition relates the interface deformation to the vertical velocity:
\begin{equation}
\partial_t h = v_y , \label{eq:h_vy}
\end{equation}
evaluated at the interface. Continuity of velocity and stresses balance at the interface impose the boundary conditions
\begin{equation}
\begin{aligned}
[\mathbf{v}]_0 &= 0 ,\\
[T_{xy}]_0 &= 0 ,\\
[T_{yy}]_0 &= -\gamma_0 \partial_x^2 h(x,t) ,
\end{aligned} \label{eq:bc}
\end{equation}
where $[s(x)]_0 \equiv s(x,0^{-}) - s(x,0^{+})$ denotes the variation of quantity $s$ across the interface, $\textbf{\textit{T}}^i=\nu_i\left(\nabla \mathbf{v}^i+\nabla{ \mathbf{v}^i}^T\right)-P^i\textbf{\textit{I}}$ are the stress tensor for each flow, $\textbf{\textit{I}}$ is the unit tensor. 
The last line of Eq.\eqref{eq:bc} expresses the balance between the normal bulk stresses and the Laplace pressure, where 
$\gamma_0$ is the interfacial tension.

Solving Eqs.\eqref{eq:nseq2}-\eqref{eq:bc} yields the dispersion relation for the interface:
\begin{equation}
i \tilde{\omega} = - \frac{[\nu_1(1+\beta_1)+\nu_2(1+\beta_2)] \gamma_0 k}{f_1 \nu_1^2 + f_2 \nu_1 \nu_2 + f_3 \nu_2^2} . \label{eq:final_dispersion}
\end{equation}
where
\begin{align*}
f_1 &= -1 + 3\beta_1 + \beta_1^2 + \beta_1^3 ,\\
f_2 &= 2 - (\beta_1 + \beta_2) + \beta_1^2 + \beta_2^2 + 4\beta_1\beta_2 + \beta_1\beta_2(\beta_1 + \beta_2) ,\\
f_3 &= -1 + 3\beta_2 + \beta_2^2 + \beta_2^3 ,
\end{align*}
and $\beta_i = \sqrt{1+\frac{i \tilde{\omega} +\alpha_i}{\nu_i k^2}}$. This resulting dispersion relation agrees quantitatively with that from numerical simulations at small wavenumber (Fig. 3{d}, Supplementary Figure~28). 
\section*{Data availability}

Source data underlying the graphs in the main figures and Supplementary Information are provided with this paper as Source Data files. Representative microscopy images and movies are provided in the main figures, Supplementary Information and Supplementary Movies.

\section*{Code availability}

The custom code used to perform the simulations and data analyses in this study has been deposited in Code Ocean and will be made publicly available upon publication.
\section*{References}
\bibliographystyle{naturemag}
\bibliography{ref}

\vspace{5mm} 

\section*{Acknowledgements}

We thank Shuo Guo for providing \textit{E. coli} BW25113 strain. Discussions with Xiaqing Shi, Hugues Chat\'e, Rui Ma, and Dahai He are highly beneficial. 

\section*{Funding Statement}
HPZ was supported by the National Natural Science Foundation of China (No. 12225410, No. 12074243) and the National Key R\&D Program of China (grant no. 2021YFA0910700). ZY was supported by the National Natural Science Foundation of China (No. 12374219), the National Key Research and Development Program of China (No. 2023YFA1407500), and the 111 project (B16029).
\section*{Author Contributions Statement}
H.P.Z. and Z.Y. conceived and supervised research. K.C. and H.P.Z. designed experiments. K.C., M.Y. and M.S. performed experiments. Y.L., K.C. and Z.Y. formulated analytical and numerical models, and performed simulations. All authors analysed data and prepared the manuscript.

\section*{Competing Interests Statement}

The authors declare no competing interests.

\end{document}